# Test adequacy evaluation for the user-database interaction: a specification-based approach

Raquel Blanco
Computer Science Department
University of Oviedo
Gijón, Spain
rblanco@uniovi.es

Javier Tuya
Computer Science Department
University of Oviedo
Gijón, Spain
tuya@uniovi.es

Rubén V. Seco
Computer Science Department
University of Oviedo
Gijón, Spain
valdesruben@uniovi.es

*Abstract—* Testing a database application is a challenging process where both the database and the user interaction have to be considered in the design of test cases. This paper describes a specification-based approach to guide the design of test inputs (both the test database and the user inputs) for a database application and to automatically evaluate the test adequacy. First, the system specification of the application is modelled: (1) the structure of the database and the user interface are represented in a single model, called Integrated Data Model (IDM), (2) the functional requirements are expressed as a set of business rules, written in terms of the IDM. Then, a MCDC-based criterion is applied over the business rules to automatically derive the situations of interest to be tested (test requirements), which guide the design of the test inputs. Finally, the adequacy of these test inputs is automatically evaluated to determine whether the test requirements are covered. The approach has been applied to the TPC-C benchmark. The results show that it allows designing test cases that are able to detect interesting faults which were located in the procedural code of the implementation.

## I. INTRODUCTION

Database applications play an important role in today's commercial systems. In these applications the business logic is usually implemented by means of a combination of imperative languages and the SQL language [15]. Testing database applications is especially important, because faults can appear in the procedural code of the program, in the SQL queries used to interact with the database, in the schema of the database or in the data stored in the database. Besides a fault may not only produce an incorrect output to the user, but also cause damage or loss of vital data for a company, which is stored in the database. So, incorrect data are also dangerous since they may be the input of other processes of the application, which may have a malfunction and produce additional damage to the data.

A typical operation of a database application, from now on called *user transaction*, starts with the selection of some data from the database to be shown in the user interface. The user, based on this information, introduces new data in this interface. Then, the transaction is executed taking into account the database state and the data supplied by the user. After that execution, the database is updated and/or new data are shown in the user interface.

A user transaction has two kinds of inputs and two kinds of outputs, which are taken from the user interface and the database. Therefore, we consider that the test input of a test case is composed of the values supplied by the user in the user interface (henceforth *user input*) along with the state of the database before the execution of the user transaction (henceforth *test database*). In the same way, the test output of a test case is formed by the values shown in the user interface (henceforth *user output*) and the state of the database after the execution of the user transaction (henceforth *output database*).

In order to guide the generation of the test inputs and to evaluate their adequacy, different criteria have been defined in the literature [33]. These criteria determine situations of interest to be tested, which are called from now on *test requirements*. Regarding to the database applications, typical criteria for procedural code have been used, such as branch coverage in the work of Emmi et al. [10], and new criteria specially designed to deal with the particularities of the source code that accesses to the database have been developed. For example, the data-flow criteria defined by Kapfhammer and Soffa [16], the structural and data-flow criteria elaborated by Willmor and Embury [26], the multiple condition coverage described by Suárez-Cabal and Tuya [21], or the SQLFpc criterion defined by Tuya et al. [24]. These approaches use the SQL statements statically defined by the application to derive the test requirements. Other works address the problem of the dynamic construction of the SQL statements to be executed in the database application and define adequacy criteria based on the number of dynamic statements covered, such as the command form coverage proposed by Halfond and Orso [11]. Recent works of Zhou and Frankl [30][31][32] also take into account the generation of the SQL statements on the fly and describe a mutation testing approach to test Java database applications, which is based on the mutation operators for SQL statements proposed by Tuya et al. [23].

However, when these criteria are used, the generation of the test inputs that cover the test requirements is guided by the implementation of the database application, instead of being led by what should be implemented according to the system specification. So the test inputs are designed to cover the structure of the source code, not to cover the expected behaviour of the application expressed in the system specification. The test cases that rely on the source code might not expose faults if the implementation does not fulfil the system specification, because what the application does, it does well, but the application does not do what it should do.

A possible approach to guide the generation of the test inputs from what the application must do consists of applying Model-based testing, whose benefits, such as its support for test automation, have been discussed in several works (for example [13]). In this approach the intended behaviour of an application is represented by means of models that are precise enough to be the basis of the derivation of meaningful test cases [25]. To achieve the goal of obtaining the test inputs of these meaningful test cases one or several models are designed from the system specification focused on testing objectives, and then a given adequacy criterion is used to derive the test inputs from these models and evaluate the adequacy achieved.

Related to the scope of testing database applications, the process of modelling the system specification involves the design of models for the required functionality (such as use cases, scenario diagrams, business rules, etc.), models for the data handled by the user, that is the user interface, (for example class diagrams, task models, etc.) and models for the data handled and stored by the application, that is the database, (for instances, relational models, a set of constraints, etc.) However, the use of different types of models, that only represent a part of the database application, complicates the tasks of deriving the test inputs for a complete view of the application and evaluating their adequacy, due to the interrelation among these models: the behaviour is affected by the data that are present in both the user interface and database, the information shown in the user interface depends on the data stored in the database and the operations that are carried out over the database also depend on the data of the user interface.

To automate the process of testing a database application from the system specification the following phases can be considered: (1) the modelling of the system specification, (2) the definition of an adequacy criterion to derive the test requirements, (3) the evaluation of the adequacy achieved by the current test inputs (user inputs and test database), (4) the generation of new test inputs to cover the test requirements that have not been covered yet, (5) the specification of the expected test outputs (user outputs and output database), (6) the execution of the database application with the test inputs and (7) the comparison between actual and expected outputs. The approach presented in this paper deals with 1, 2 and 3: the modelling of the system specification, the definition of an adequacy criterion over the model defined to derive the test requirements and the automatic evaluation of the adequacy achieved by the test inputs generated.

The main contributions of this work are:
- The definition of an Integrated Data Model, where the two types of inputs of a database application, that is the user interface and the database, are modelled in a unified way.
- The modelling of the required functionality of the database application through a set of business rules that take into account the Integrated Data Model defined.
- The elaboration of a MCDC-based criterion over the business rules to automatically derive the test requirements.
- The automation of the evaluation of the test inputs adequacy that involves both user inputs and test database.

The remainder of the paper is organized as follows: Section II describes our approach to automate the generation of the test requirements from the system specification and the evaluation of the test inputs adequacy. Section III presents the results of the experiments over a case study. Section IV presents the related work. The paper ends with conclusions and future work.

## II. PROBLEM APPROACH

To describe the problem approach let us consider a user transaction called "new-order transaction", taken from the TPC-C benchmark [22]. This benchmark represents the activity of a wholesale supplier that has several sale districts and associated warehouses, which have stocks for a number of items. Each customer of the wholesale supplier company is served by a specific warehouse. When a customer places an order in the user interface of the user transaction, all items are supplied by the associate warehouse (by default), but if an item is not in stock of this warehouse, the customer must indicate a different warehouse to get the item. To calculate the total price of an order, the warehouses used to serve the items are considered and, as a result, the price when all items are ordered to the customer's warehouse is different to the price when some items are ordered to an alternative warehouse.

According to this example, some interesting test requirements are: (1) all items are ordered to the customer's warehouse because all of them are in stock, (2) some items are ordered to a different warehouse because they are not in stock of the customer's warehouse. The test database has to incorporate meaningful data to cover these situations and the user has to introduce the user inputs: for (1) the user has to supply a customer and a list of items such that in the test database all of them are in stock of the customer's warehouse and for (2) the list of items supplied by the user must contain at least one item that in the test database is not in stock of the customer's warehouse but is in stock of another warehouse.

Our approach considers this user transaction as a unit to be tested. Each test unit is called from now on *test assignment*. For each test assignment several test cases are designed to cover its test requirements, and each of them is composed of different user inputs and usually the same test database (to reduce the cost of test preparation and execution).

As stated above, the system specification of a test assignment includes the description of its required functionality, the structure of the information handled and stored by the application (database) and the structure of the information handled by the user (user interface). These three parts are considered together to evaluate the adequacy of the test inputs.

The user input and the test database that compose a test input are closely related, as the previous example shows. For instance, to achieve the second test requirement, the customers, warehouses and items in stock of each warehouse that are stored in the database must be considered to introduce in the user interface (1) a customer that is served by a warehouse with items in stock, (2) one or several items that are in stock of this warehouse and (3) one or several items that are in stock of a different warehouse and are not in the stock of the customer's warehouse. To represent both the user input and the test database and their dependences in a homogeneous way, our approach integrates both user interface and database into a unique model called Integrated Data Model (IDM), which is described in Section II.A. The IDM contains the structure of the test cases designed for the test assignment.

On the other hand, our approach represents the required functionality of a test assignment as a set of business rules from which the test requirements can be obtained automatically (a business rule is a statement that defines or constrains the business structure or the business behaviour [12]). The description of this functionality involves the definition of the properties that both the information stored in the database and handled by the user must fulfil, the actions that are carried out over this information and the output of these actions. As both user interface and database are represented by the IDM, the business rules express the functionality of the test assignment in terms of the IDM, using the language presented in Section II.B. By means of the IDM the process of evaluating the adequacy of the test inputs is simplified, since it is carried out as if the test assignment only had one type of input.

Our approach performs the following steps, which are depicted in Figure 1:
- *Step 1*: the model IDM that integrates both user interface and database is created to represent the test inputs.
- *Step 2*: the required functionality of the test assignment is represented as a set of business rules that are expressed in terms of the IDM.
- *Step 3*: the test requirements are derived from the business rules using a MCDC-based criterion and they are automatically evaluated.
- *Step 4*: the test inputs designed, taking into account the IDM, are automatically evaluated to determine the test requirements covered.
- *Step 5*: if some test requirements have not been covered yet, new test inputs are designed (the automation of this process is out of the scope of this paper) and then a new evaluation of the test inputs adequacy is carried out.

To automatically apply the adequacy criterion and check whether the test inputs cover the test requirements, we express the IDM as a relational model and generate SQL queries as an executable representation of the test requirements. So the SQL queries are executed against a database derived from the IDM to evaluate whether the data stored cover the test requirements.

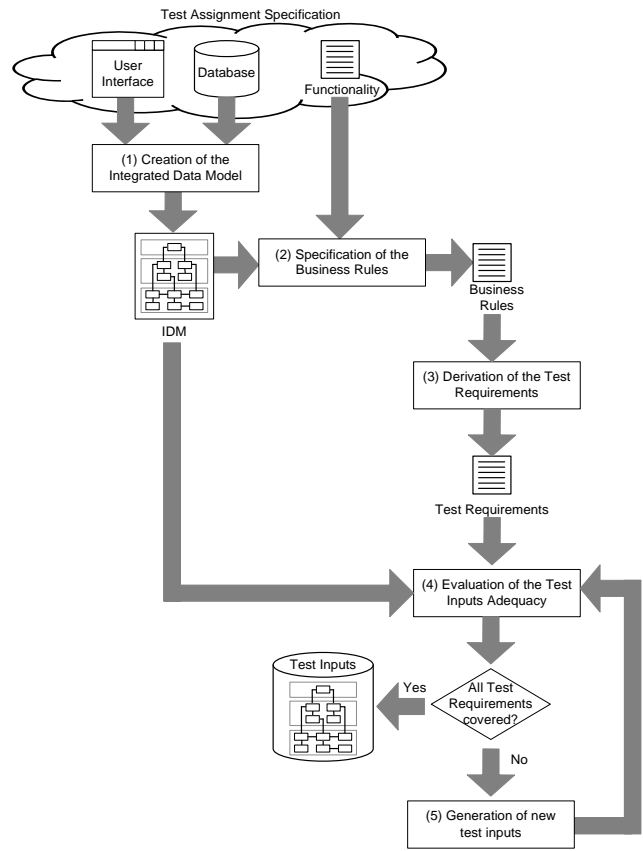

Figure 1. General Schema

### A. Integrated Data Model

To cover the test requirements derived from the adequacy criterion used in the testing process of a database application a set of test cases is designed. Each test case involves the introduction of the user input in the user interface and the population of a test database. To reduce the cost of populating the database, it is useful to share the same test database by the most number of test cases as possible, which may have been created for different test assignments, so the test inputs of these test cases only differ in the user inputs. To represent these test inputs our approach defines the Integrated Data Model (IDM), where the user interfaces of the test assignments and the database used are integrated into a unique model and the state of the test database can be shared by several test cases.

The IDM is composed of three levels, as it is shown in Figure 2:
- *Database level*, which models the database used by the test assignments of the database application and represents the test database.

- *UI level*, which models the user interfaces of the test assignments and represents the user input of each test case. This level is connected to the Database level as the user inputs are closely related to the values stored in the test database.
- *Test Case level*, which represents the test cases created for the test assignments. These test cases are related to different user inputs and share the test database. This level is connected to the UI level to identify the user input that corresponds to each test case.

Each level is modelled through a relational model composed of a set of entities and their relationships, called *intra-level relationships*. The connections between different levels are relationships, which are called *inter-level relationships*, among two entities that belong to these different levels.

To illustrate the creation of the Integrated Data Model, consider again the test assignment "new-order transaction" of the TPC-C benchmark. Figure 3 shows its user interface, which is composed of the general information of an order (such as the customer and the discount and warehouse for this customer) and the list of items ordered along with their information (such as the item id, the quantity ordered and the item price). Each field of the user interface has a superscript which indicates whether it is an input variable (superscript I) or an output variable (superscript O).

The IDM of this example is depicted in Figure 4. The database level is formed by the relational model of the database used by all transactions of the TPC-C benchmark. The UI level contains the model for the "new-order transaction" (the models derived from the other test assignments of the TPC-C benchmark are not represented in the figure). The name of each entity of the UI level starts with the prefix UI to identify the entities that clearly belong to this level.

The user interface of this test assignment is represented by two entities: UI_Order, which represents the general information of an order, and UI_OrderLine, which represents each item ordered. There is an intra-level relationship between these entities which indicates that an order is composed of a list of items.

The connections between the UI level and the database level indicate that the data in both user interface and database are related. Thus, the entity UI_Order has an inter-level relationship with the entity Customer and another one with the entity Order. The former indicates that an order of a user input, represented by UI_Order, belongs to a specific customer stored in the test database, which can be used in several test cases. The latter relates the order of a user input with the order stored in the database level after executing the test case. The entity UI_OrderLine has an inter-level relationship with the entity Stock, where the items in stock of each warehouse are stored. This relationship specifies that an item ordered in the user input is in the stock of a warehouse in the test database. An item of this stock can be ordered more than once.

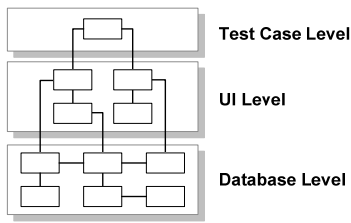

Figure 2. Structure of the Integrated Data Model (IDM)

Figure 3. Inputs of the test assignment "new-order transaction"

Finally, the Test Case level is composed of an entity called TestCase where each test case is distinguished. The entity Test Case has an inter-level relationship with the test assignment in the UI level to indicate that several test cases can be designed for the test assignment and each of them corresponds to a specific order introduced in the user interface (the user input).

The IDM is physically implemented as a database, where the structure and the value of the test inputs are stored. Each test case is represented by a specific tuple in the entity

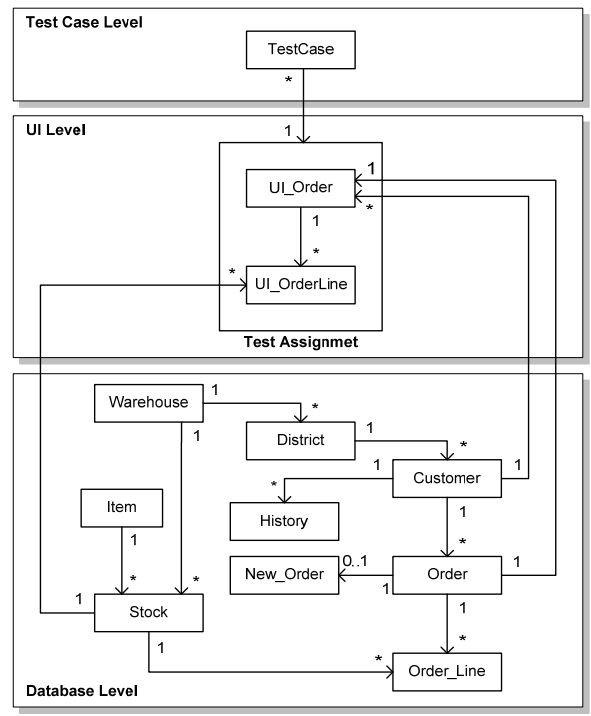

Figure 4. IDM of the test assignment "new-order transaction"

TestCase. The user input of each test case is the set of tuples stored in the entities of the UI level that are related to the tuple of TestCase used to identify it. The test database is composed of the tuples stored in the entities of the Database level.

### B. Specification of business rules

This section describes how the business rules are constructed from the specification of the test assignments. Two kinds of rules are considered: constraint and derivation business rules [12]. The constraint business rules impose conditions on the state of the entities of the IDM. The derivation rules infer new knowledge from the state of these entities. The vocabulary used in their construction is based on the SBVR specification [20].

Before describing the business rules, it is necessary to define some concepts that are used in their construction: paths, path attributes, frames and frame attributes. In these definitions the notation presented by Codd [7] is used.

The entities referred by a business rule can be related by several paths over the IDM and, therefore, it is necessary to determine which one is used in order to identify the relationships involved:

*Definition 1*: a *path* P is a sequence of one or more entities $R_1, R_2, \ldots, R_n$ of the IDM, where each pair $(R_i, R_{i+1})$ is directly connected via some attributes in the predicate $q_{i,i+1}()$:

path P is $R_1[q_{1,2}()]R_2[q_{2,3}()]\ldots[q_{n-1,n}()]R_n$

If a pair $(R_i, R_{i+1})$ is connected via the foreign key, it is not necessary to specify the predicate $q_{i,i+1}()$.

A path P establishes the context on which the conditions of the business rules are applied. This context is composed of the tuples obtained from the Cartesian product of the entities $R_i$ (i=1..n) that fulfil the predicates $q_{i,i+1}()$.

*Definition 2*: a *path attribute* is an attribute A of a path P denoted by P.A. If A is not unique in P it is denoted by P.R.A, where R is an entity of P that contains A.

*Definition 3*: a *frame* G is a set of tuples over a path P which have the same value for one or more path attributes $A_1, A_2, \ldots, A_n$:

frame G is P///$A_1, A_2, \ldots, A_n$

*Definition 4*: let G be a frame over a path P. A *frame attribute* is a path attribute A of P whose value is analyzed in the tuples of the context defined by G. The frame attribute is denoted by G.A. If A is not unique in P it is denoted by G.R.A where R is an entity of P that contains A.

The business rules establish conditions over the values of path attributes and frame attributes, called from now on *value conditions*, and conditions regarding the number of tuples of a path that relates two entities, called from now on *quantification conditions*. The definitions of both types of conditions are presented below, using the EBNF notation [14].

*Definition 5*: a *value condition* defines or constraints the value of a path attribute P.A (or a frame attribute G.A):

value_condition = simple_condition | range_condition
simple_condition = (P.A | G.A) (at least | at most | exactly
    | different to | like) $q$
range_condition = (P.A | G.A) at least $p$ and at most $q$

where $p$ and $q$ are arithmetic expressions over path attributes of P or frame attributes of G or constants.

*Definition 6*: a *quantification condition* defines or constraints the number of tuples of an entity P.S that are related to a specific tuple of an entity P.R:

quantification_condition = simple_quantification |
    range_quantification
simple_quantification = P.R (at least | at most | exactly
    | different to) $q$ P.S
range_quantification = P.R at least $p$ and at most $q$ P.S

where $p$ and $q$ are expressions whose evaluation returns an integer number greater than or equal to 0.

The following subsections present the patterns that allow the construction of different business rules using the EBNF notation (Section 1 describes the constraint business rules and Section 2 describes the derivation business rules). Each business rule is illustrated through an example based on the specification of the TPC-C benchmark.

#### 1) Constraint business rules

A constraint business rule establishes one or several value or quantification conditions to be fulfilled by each tuple or a set of tuples of the path that defines the context for the business rule. Taking into account the type of these conditions, two kinds of rules are defined: *constraint rule for values* and *constraint rule for the number of tuples in a path*.

*Definition 7*: The general pattern of a constraint business rule is:

each $c_i$ {(and | or) each $c_i$}

where $c_i$ is a condition written according to the patterns defined by each type of constraint rule (see definitions 8 and 9 below).

*Definition 8*: a *constraint rule for values* is a constraint business rule that establishes one or several value conditions over the path attributes $P.A_i$ (i=1..n) that must be fulfilled by each tuple of P. Each $c_i$ of the general pattern of a constraint rule is defined as:

$P.A_i$ must be (at least $p$ | at most $p$ | exactly $p$ |
    different to $p$ | like $p$ | at least $p$ and at most $q$)

where $p$ and $q$ are arithmetic expressions over path attributes of P or constants.

*Example 1*: consider that the quantity ordered for each item of an order introduced in the user interface must be in the range [1, 10]. The business rule is:

Each UI_OrderLine.ol_quantity must be at least 1 and at most 10

*Definition 9*: a *constraint rule for the number of tuples in a path* is a constraint business rule that establishes one or several quantification conditions, where each one restricts the number of tuples of an entity $S_i$ (i=1..n) of a path P (denoted as $P.S_i$) that can be related to a specific tuple of an entity $R_i$ (i=1..n) of P (denoted as $P.R_i$). Each $c_i$ of the general pattern of a constraint rule is defined as:

$P.R_i$ must have (at least $p$ | at most $p$ | exactly $p$ |
    different to $p$ | at least $p$ and at most $q$) $P.S_i$

where $p$ and $q$ are expressions whose evaluation returns an integer number greater than or equal to 0.

*Example 2*: consider that each order introduced in the user interface must have at least 5 and at most 15 order lines. The business rule is:

    Path P1 is UI_Order[]UI_OrderLine
    Each P1.UI_Order must have at least 5 and at most 15
        P1.UI_OrderLine

*2) Derivation business rules*

A derivation business rule infers new knowledge when one or several value or quantification conditions are true in some or a group of tuples of the path that defines the context for the business rule.

Taking into account the type of conditions and how many tuples of the path must fulfil the conditions, the following rules are defined: *derivation rule for conditions in some tuple in a path*, *derivation rule for conditions in all tuples of a frame* and *derivation rule for the number of tuples in a path*.

*Definition 10*: The general pattern of a derivation rule is:
    if $c_i$ {(and | or) $cond_i$} then *r*

where $c_i$ is a condition written according to the patterns defined by each type of derivation rule (see definitions 11 to 13 below) and *r* is a set of actions used to infer the new information.

*Definition 11*: a *derivation rule for conditions in some tuple in a path* is a derivation business rule that establishes one or several value conditions over the path attributes P.$A_i$ (i=1..n) to be fulfilled by some tuple of P. Each $c_i$ of the general pattern of a derivation rule is defined as:
    P.$A_i$ is (at least *p* | at most *p* | exactly *p* | different to *p* | like *p* |
        at least *p* and at most *q*)

where *p* and *q* are arithmetic expressions over path attributes of P or constants.

*Example 3*: consider that the brand information of an item of an order introduced in the user interface is inferred when the brand information stored for that item in both entities Stock (attribute s_data) and Item (attribute i_data) include the string 'ORIGINAL'. The business rule is:
    Path P2 is UI_OrderLine[]Stock[]Item
    If P2.i_data is like '%ORIGINAL%' and P2.s_data is
        like '%ORIGINAL%' then P2.o_brand = 'B'

*Definition 12*: a *derivation rule for conditions in all tuples of a frame* is a derivation business rule that establishes one or several value conditions over the frame attributes G.$A_i$ (i=1..n) to be fulfilled by all tuples of G. Each $c_i$ of the general pattern of a derivation rule is defined as:
    each G.$A_i$ is (at least *p* | at most *p* | exactly *p*| different to *p* |
        like *p* | at least *p* and at most *q*)

where *p* and *q* are arithmetic expressions over frame attributes of G or constants.

*Example 4:* consider that the attribute o_all_local of an order, which is been stored in the database, is inferred when all items of this order introduced in the user interface are supplied by the warehouse that servers the customer.
    Path P3 is UI_OrderLine[]UI_Order[]Order
    Frame G is P3///o_tc_id, o_ui_id
    If each G.ol_supply_w_id is exactly G.o_w_id then
        G.o_all_local = 1

The path P3 is framed according to the derived attributes that identify each order introduced in the user interface. When all tuples of a frame fulfil the condition defined over the frame attributes, the new knowledge is inferred for this frame, in this case for the order.

*Definition 13*: a *derivation rule for the number of tuples in a path* is a derivation business rule that establishes one or several quantification conditions to define the number of tuples of an entity P.$S_i$ (i=1..n) that must be related to a specific tuple of an entity P.$R_i$ (i=1..n). Each $c_i$ of the general pattern of a derivation rule is defined as:
    P.$R_i$ has (at least *p* | at most *p* | exactly *p* | different to *p* |
        at least *p* and at most *q*) P.$S_i$

where *p* and *q* are expressions whose evaluation returns an integer number greater than or equal to 0.

*Example 5*: consider that the status of an order introduced in the user interface is updated with the value 'error' when the order has more than 15 order lines. The business rule is:
    Path P4 is UI_Order[]UI_OrderLine
    If P4.UI_Order has at least 16 P4.UI_OrderLine then
        P4.Status = 'error'

*C. Coverage Evaluation*

To obtain the set of test requirements from the business rules, a Masking MCDC-based criterion is used. This criterion requires that every condition in a decision in the program has taken on all possible outcomes at least once, every decision in the program has taken all possible outcomes at least once, and each condition in a decision has been shown to independently affect the decision's outcome [6].

In our case, the program is represented by means of a set of business rules, where each one establishes a set of conditions that forms the decision on which the Masking MCDC-based criterion is applied. On the other hand, the evaluation of the decision and the conditions of a business rule is affected by the context on which they are considered. As stated above, this context depends on the predicates of the path P used in the business rule. So, our approach also applies the Masking MCDC-based criterion over the predicates of P to derive different contexts on which the conditions of each business rule are found to hold true.

The process of automatically deriving the test requirements and evaluating the coverage achieved by the test inputs has been implemented by a set of tools (see Figure 5).

First, the model of the user interface of each test assignment is represented through a SQL-like language. Then, the UIBDRules tool generates the database that represents the IDM, called henceforth *IDM database*, taking as input the database schema of the application and the representation of the user interface. As stated above this database contains the structure of the test cases and for each one it stores the test input.

Next, UIBDRules derives the test requirements from each business rule in the form of SQL queries, which can be executed against the IDM database. To do this, our approach relies on a Masking MCDC-based criterion specially tailored to deal with SQL, called SQLFpc [24]. This criterion is applied over a SQL query and derives its test requirements as SQL queries, called *coverage rules*. The generation of the coverage rules has been implemented in the SQLFpcWS web service [24]. In other words, UIBDRules transforms each business rule into one or several SQL queries and then invokes SQLFpcWS with each query and the schema of the IDM database to obtain the coverage rules. After that, UIBDRules modifies and filters some coverage rules to get the final set of coverage rules that constitutes the executable representation of the test requirements of the business rule.

The final set of coverage rules is presented, along with a description in natural language, to the tester, who populates the IDM database with the test inputs intended to fulfil the conditions imposed by the coverage rules. Afterwards, each coverage rule is executed against the IDM database to determine whether it is covered by the test inputs stored, that is, whether at least one tuple is obtained after its execution. Thus, the SQL queries guide the design of the test inputs and the execution of these SQL queries against the IDM database allows the automatic evaluation of the test inputs adequacy.

To illustrate the test requirements derived from a business rule using the Masking MCDC-based criterion, and their executable representation, let us consider the business rule of *Example 3*. For this business rule, UIBDRules generates coverage rules that represent the test requirements derived from the path P2 (1 to 3) and the test requirements derived from the conditions established by the business rule (3 to 7):

(1) The predicate of P2 that connects UI_OrderLine and Stock is not fulfilled and the other predicate of P2 is fulfilled. Besides, all conditions of the business rule are found to hold true.
(2) The predicate of P2 that connects Stock and Item is not fulfilled and the other predicate of P2 is fulfilled. Besides, all conditions of the business rule are found to hold true.
(3) All predicates of the path P2 are fulfilled and all conditions of the business rule are found to hold true.
(4) The condition *P2.i_data is like '%ORIGINAL'* is true and the condition *P2.s_data is like '%ORIGINAL'* is false. Besides, all predicates of P2 are fulfilled.

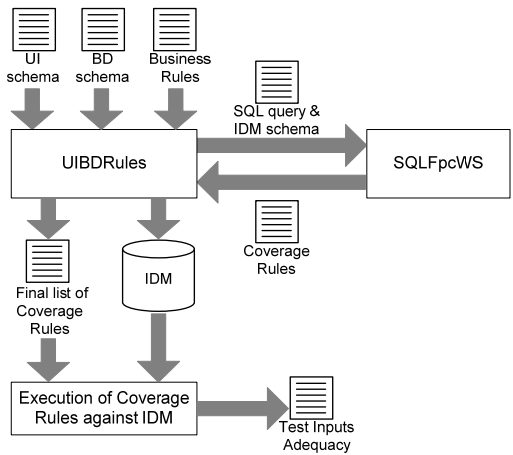

Figure 5. Implementation schema

(5) The condition *P2.s_data is like '%ORIGINAL'* is true and the condition *P2.i_data is like '%ORIGINAL'* is false. Besides, all predicates of P2 are fulfilled.
(6) The path attribute P2.i_data has a missing value and the condition *P2.s_data is like '%ORIGINAL'* is true. Besides, all predicates of P2 are fulfilled.
(7) The path attribute P2.s_data has a missing value and the condition *P2.i_data is like '%ORIGINAL'* is true. Besides, all predicates of P are fulfilled.

The test requirements 6 and 7 take into consideration that the attributes of the entities of the database used by the application and also the input variables of the user interface can have a missing value when the test assignment is executed.

The following coverage rule constitutes the executable representation of the test requirement 4:

```
SELECT *
FROM TestCase
INNER JOIN UI_Order ON (tc_id = o_tc_id)
INNER JOIN UI_OrderLine ON (o_tc_id = ol_tc_id
        AND o_ui_id = ol_ui_id)
INNER JOIN Stock ON (s_i_id = ol_i_id
        AND s_w_id = ol_supply_w_id)
INNER JOIN Item ON s_i_id = i_id
WHERE NOT(s_data like '%ORIGINAL%')
        AND (i_data like '%ORIGINAL%')
```

The INNER JOIN clauses among the entities that form the path P2 represent the fulfillment of the predicates of P2. The WHERE clause expresses the conditions over the path attributes P2.i_data and P2.s_data. Besides, additional INNER JOIN clauses are introduced to relate, through the foreign keys, the entities of P2 with the entity TestCase, where each test case designed is identified. In this case, the entity UI_OrderLine is related to TestCase through UI_Order. Thus, the test input inserted in the IDM database to cover the previous SQL query is associated with a specific test case.

To cover this coverage rule, the IDM database must contain tuples in the entities TestCase, UI_Order, UI_OrderLine, Stock and Item that fulfil the predicates of the INNER JOIN and WHERE clauses. Furthermore, the referential integrity with the entities Customer, District and Warehouse must be fulfilled.

III. CASE STUDY

In this section, a case study is presented using the standard specification of the TPC-C benchmark [22] as the system under test. This benchmark defines five user transactions (each one constitutes a test assignment): *New-Order*, *Payment*, *Delivery*, *Order-Status* and *Stock-Level* (both New-Order and Payment transactions are mid-weight read-write transactions and the others are mid-heavy-weight read transactions, as they are defined in [22]).

Among the different implementations for this benchmark, the open source benchmark called BenchmarkSQL [1] was selected, which closely resembles the TPC-C standard for OLTP. This implementation has been adapted so that the emulated user takes the input data from the IDM. The implementation has 129 decisions and 35 SQL queries.

The specification of the TPC-C benchmark was analyzed to obtain a set of business rules, then the Masking MCDC-based criterion was applied and the executable representation of each test requirement was obtained, that it, the coverage rules. The IDM was implemented in an Oracle database. The test cases were generated by inspecting each coverage rule and then filling the required tuples in order to cover it. A unique IDM database was used to store the test cases of all test assignments, which share the Database level. First, the IDM database was empty and then it was incrementally populated with the tuples that covered the uncovered coverage rules of each test assignment. The test assignments were analyzed in the following order: New-Order, Payment, Order-Status, Delivery and Stock-Level.

Table 1 displays, for each test assignment, the number of business rules and the number of coverage rules generated. From the 26 business rules, 9 are constraint rules for values, 1 is a constraint rule for the number of tuples in a path, 9 are derivation rules for conditions in some tuple in a path, 1 is a derivation rule for conditions in all tuples of a frame and 6 are derivation rules for the number of tuples in a path. Additionally, Table 1 displays the total number of tuples inserted into the IDM database, which accumulates the number of tuples inserted into the Test Case level (one per test case), into the UI level and into the Database level. As the Database level is shared by all test assignments, the column "Database level" indicates the number of tuples inserted in that level to cover the coverage rules that had not been covered yet by the exiting tuples.

Then, the test cases were executed against BenchmarkSQL and all detected failures recorded. At present the expected output is specified by the tester. For each test assignment, Table 1 also shows the number of test cases, the total number of failures found and the number of failures after removing duplicates. In total, 13 non duplicate failures were detected.

Finally, the source code of BenchmarkSQL was analyzed to find the defects that caused the failures. All of them have been found in the procedural code, instead of the SQL queries.

From the total number of failures, 5 are caused by faults related to input validation errors, such as the creation of an order in the user interface with more order lines than those allowed or the specification of a threshold for the stock level comparison out of the allowed range. The other 8 failures correspond to faults in the implementation of BenchmarkSQL: 3 failures are caused by an incorrect handling of null values of the input data that may be present in the test database, 3 failures are caused by an incorrect update of the user interface and 2 failures are derived from the incorrect processing of the test inputs. To illustrate the kind of defects that were found, some representative faults are described below:

- Null values: the specification of the Payment transaction indicates that when a customer is making a payment, its attribute c_credit is analyzed and only if it has the value 'BC' another attribute called c_data is updated. When a null value of c_credit is processed, it is not checked and then an exception is thrown.
- Incorrect update of the user interface: considering the same specification of the previous example, the application correctly updates the attribute c_data in the database when c_credit is 'BC'. However, the user interface is not correctly updated, since the application uses a variable that does not take into account the output database state.
- Incorrect processing of test inputs: the specification of the New-Order transaction indicates how to infer the brand information shown in the user interface for an item of an order. The brand information has the value 'B' when two attributes of the database (Stock.s_data and Item.i_data) related to the item include the string 'ORIGINAL', and otherwise the value shown is 'G'. However, the implementation uses the string 'GENERIC' in the comparison, and the output given to the user is incorrect when the value of both attributes includes 'ORIGINAL'.

IV. RELATED WORK

There exist other approaches in the literature which address the problem of testing database applications, taking into account the database and the user input. Chays et al. [3][4] describe the AGENDA tool, which has been improved in [5][9]. The tool takes as input the database schema, the application source code (which consists of a set of SQL queries), suggested values for the attributes given by the user and some test heuristics. With this information, the tool populates the database and generates values for the input variables of the application that are present in the SQL queries. Although our approach does not automate the generation of the test inputs, the derivation of the test requirements relies on a more complete specification of the database application. The use of a Masking MCDC-based criterion allows obtaining a set of meaningful test inputs that checks both the equivalence classes obtained from the conditions of the business rules and their decisions.

Emmi et al. [10] propose an algorithm that generates input data for the program and database states to cover all branches of the procedural code, taking into account that the execution of a branch may depend on the outcome of a SQL query. In their work the test requirements are derived from the conditional statements of the procedural code and from the WHERE clause of each SQL query whose outcome affects the coverage of a branch in the procedural code. However, our approach derives the test requirements from the conditions defined in the system specification, which can be implemented correctly (or incorrectly) in the procedural code and/or in the SQL queries embedded. This specification-based approach can guide the generation of meaningful test inputs that complement those obtained by the work of Emmi et al.

Table 1. Results obtained for each test assignment

| Test Assignment | Business Rules | Coverage Rules | Tuples Inserted | | | Test Cases | Failures | Non Duplicate Failures |
| | | | IDM | UI Level | Database Level | | | |
|---|---|---|---|---|---|---|---|---|
| New-Order | 7 | 38 | 113 | 86 | 17 | 10 | 10 | 5 |
| Payment | 6 | 24 | 23 | 9 | 5 | 9 | 7 | 5 |
| Order-Status | 5 | 19 | 30 | 7 | 16 | 7 | 0 | 0 |
| Delivery | 4 | 20 | 27 | 7 | 13 | 7 | 4 | 2 |
| Stock-Level | 4 | 24 | 129 | 8 | 113 | 8 | 3 | 1 |
| **Total:** | **26** | **125** | **322** | **117** | **164** | **41** | **24** | **13** |

Zhou and Frankl [30][31][32] describe the JDAMA tool, which applies mutation testing over the SQL queries embedded into a Java program. These queries can contain input variables of the program and are dynamically generated. As the mutants are only obtained from the SQL queries their approach relies on the extensive use of SQL queries in the source code to obtain a meaningful set of test inputs for the database application. In contrast, our approach does not depend on the use of SQL queries in the source code of the database application to guide the creation of the test inputs.

The closest approach to ours is that of Willmor and Embury [28] who applied their intensional approach of [27] to verify the implementation of constraint business rules. Each test case incorporates check-conditions, which represent the business rule, pre-conditions, which are used to prepare the database for the test case execution, and post-conditions, which verify whether the execution of the test case violates the business rule. Unlike the work of Willmor and Embury, our approach uses the business rules to derive the test requirements that guide the generation of meaningful test cases for database applications, and it is able to handle the input variables provided by the user interface.

The approach presented in this paper does not address the automatic generation of test inputs. However it represents the structure of the test inputs as a database and defines the test requirements to design them as SQL queries. Thus, a number of complementary approaches can be used to automatically generate the test inputs, such as those present below.

Zhang et al. [29] describe a fault-based approach which generates a set of constraints that feeds a general purpose constraint solver to derive the test database. Binnig et al. [2] propose the Multi-RQP technique to populate a test database, taking as input a set of queries and their expected results. Khalek et al. [17] present the ADUSA tool, which automates the test database generation for a given SQL query and a given database schema using the Alloy Analyzer. A recent work of Khalek and Khurshid [18] relies on ADUSA to test DBMS engines. Lo et al. [19] describe the generator QAGen which generates a test database from an individual query and the database schema. QAGen uses symbolic query processing to capture the user-defined constraints on the query, such as the output cardinality, into the database and a constraint solver to instantiate the database. De la Riva et al. [8] propose an approach for the automatic generation of a test database for a set of SQL queries using the SQLFpc adequacy criterion. This approach models both the schema and the test requirements obtained from the adequacy criterion in the Alloy language and then the Alloy Analyzer generates the test database.

To complement our approach with the automatic generation of the test inputs (user inputs and test database), the aforementioned works could populate the IDM database. To do this, the SQL queries that represent the test requirements derived from the business rules could be translated into the constraints accepted by these works.

In general, most of the existing approaches either do not consider the user interaction or are implementation based. However, our approach takes into account the user interaction to generate the test inputs and it is specification based, so it does not depend on a particular implementation.

V. CONCLUSIONS AND FUTURE WORK

This work presents an approach to automatically derive the test requirements from the specification of a database application and to automate the evaluation of the test inputs adequacy, taking into account the database and the user interface.

The database and the user interface are integrated into a unique model called IDM that contains the structure of the test cases for the database application. The required functionality of the database application is expressed through a set of business rules, written in terms of the IDM, on which a Masking MCDC-based criterion is applied to automatically derive and evaluate whether the test requirements are fulfilled by a given set of test inputs (user inputs and test database). Thus, our approach helps the tester in identifying interesting test situations both for the user inputs and the database state.

To automate the evaluation of the test inputs adequacy, the IDM is implemented as a database, which stores both the user inputs and the test database of the test cases designed, and the test requirements are expressed as SQL queries, which are executed against the database. The execution of these queries determines the test requirements that are covered by the test inputs.

The results of the case study show that the test cases obtained are able to detect interesting faults which were located in the procedural code of an implementation of the benchmark TPC-C.

Future work includes several avenues. On the one hand, to improve the expressiveness of the business rules that can be handled, and to model more complex user transactions involving more than one interaction between the user and the application. Furthermore, to automate the population of the IDM (that is, to generate the test database and the user inputs) using, probably, some of the aforementioned approaches for test database generation. Additionally, a comprehensive experimentation should be performed to compare our approach with other works, in order to evaluate their effectiveness at detecting faults in database applications.

Finally, as the user inputs and test database are represented by the IDM in a unified way, the outputs are represented in it and then they can also be used to partially automate the comparison between actual and expected outputs. In addition, as some business rules express the output behaviour they may also be used as a test oracle.

ACKNOWLEDGMENT

This work has been funded by the Department of Science and Innovation (Spain) and ERDF funds (TIN2010-20057-C03-01).